\documentclass[12pt,twoside]{article}   
\usepackage[super,sort,comma]{natbib}

\usepackage{fancyhdr}		

\usepackage[section]{placeins}   %
\usepackage{multirow}
\usepackage{graphicx}

\makeatletter \renewcommand\@biblabel[1]{$^{#1}$} \makeatother
\newcommand{\cen}[1]{\begin{center} #1 \end{center}}

\usepackage{hyperref}
\hypersetup{ colorlinks,
    citecolor=blue,
    filecolor=blue,
    linkcolor=blue,
    urlcolor=blue
}

\usepackage{xcolor}

\definecolor{gray}{rgb}{0.6,0.6,0.6}
\definecolor{red}{rgb}{0.85,0,0}
\definecolor{green}{rgb}{0,0.85,0}
\definecolor{blue}{rgb}{0,0,0.85}
\definecolor{beige}{rgb}{0.92,0.87,0.78}
\usepackage[all]{hypcap}    

\begin{document}

\cen{\sf {\Large {\bfseries Joint enhancement of automatic chest X-ray diagnosis and radiological gaze prediction with multi-stage cooperative learning} \\  
Zirui Qiu$^1$, Hassan Rivaz $^2$, Yiming Xiao$^1$} \\
$^1$Department of Computer Science and Software Engineering, Concordia University, Montreal, Canada

$^2$Department of Electrical and Computer Engineering, Concordia University, Montreal, Canada
}

\pagenumbering{roman}
\setcounter{page}{1}
\pagestyle{plain}
Corresponding authors: leoqiuzirui@gmail.com, yiming.xiao@concordia.ca \\

\begin{abstract}
\noindent {\bf Background:} As visual inspection is an inherent process during radiological screening, the associated eye gaze data can provide valuable insights into relevant clinical decision processes and facilitate computer-assisted diagnosis. However, the relevant techniques are still under-explored.

\noindent {\bf Purpose:} With deep learning becoming the state-of-the-art for computer-assisted diagnosis, integrating human behavior, such as eye gaze data, into these systems is instrumental to help guide machine predictions with clinical diagnostic criteria, thus enhancing the quality of automatic radiological diagnosis. In addition, the ability to predict a radiologist's gaze saliency from a clinical scan along with the automatic diagnostic result could be instrumental for the end users.

\noindent {\bf Methods:} We propose a novel deep learning framework for joint disease diagnosis and prediction of corresponding radiological gaze saliency maps for chest X-ray scans. Specifically, we introduce a new dual-encoder multi-task UNet, which leverages both a DenseNet201 backbone and a Residual and Squeeze-and-Excitation block-based encoder to extract diverse features for visual saliency map prediction, and a multi-scale feature-fusion classifier to perform disease classification. To tackle the issue of asynchronous training schedules of individual tasks in multi-task learning, we propose a multi-stage cooperative learning strategy, with contrastive learning for feature encoder pretraining to boost performance.

\noindent {\bf Results:} Our proposed method is shown to significantly outperform existing techniques for chest radiography diagnosis (AUC = 0.93) and the quality of visual saliency map prediction (correlation coefficient = 0.58).

\noindent {\bf Conclusion:} Benefiting from the proposed multi-task multi-stage cooperative learning, our technique demonstrates the benefit of integrating clinicians' eye gaze into radiological AI systems to boost performance and potentially explainability.


\end{abstract}

\newpage     

\newpage

\setlength{\baselineskip}{0.7cm}      

\pagenumbering{arabic}
\setcounter{page}{1}
\pagestyle{fancy}
\section{Introduction}
The continuous advancement of deep learning (DL) in medical imaging has led to a new era in diagnostic radiology, offering diagnostic tools with unprecedented accuracy and efficiency.  In the realm of chest X-ray (CXR) analysis, where the imaging modality has poor soft tissue contrast, DL  models have shown significant potential and efficiency in identifying and classifying various pulmonary conditions \cite{goyal2023detection, lakhani2017deep}. However, the task-driven feature learning in DL models, even with various attention mechanisms \cite{oktay2018attention, rasoulian2023weakly} may not align with the clinical decision-making criteria \cite{qiu2023visual}, potentially limiting their performance and credence. Eye gaze data, which captures the inherent visual attention of clinicians during radiological reading could help address this issue by providing a unique venue to integrate human behavioral insights into medical AI systems \cite{luis2023integrating}. Recently, several DL models \cite{kato2022classification, wang2023gazegnn} have been proposed to leverage recorded gaze data as an input or as an auxiliary task to enhance diagnostic accuracy. As gaze data collection requires specialized devices and postprocessing, the latter strategy in a multi-task learning framework appears more attractive in practice. Although multi-task learning could potentially boost the performance of individual tasks in synergy \cite{zhu2022multi}, the loss of each task can overfit at different rates if care is not taken, and may restrict their performance. 

To address the aforementioned challenges, we introduce a novel multi-task DL model based on a cooperative learning strategy to enhance disease classification performance and clinical visual saliency map prediction simultaneously for chest X-ray. Our contributions are threefold: \textbf{First}, to mitigate the issue of asynchronous training schedules of individual tasks, we proposed a multi-stage cooperative learning strategy, which helps with robust training of the designed multi-task network. \textbf{Second}, we designed a novel dual-encoder UNet with multi-scale feature-fusion to enhance multi-task collaboration for optimal outcomes. \textbf{Lastly}, our experiments demonstrated the superior performance of the proposed method than existing methods, and we share our code publicly at \textit{https://github.com/HealthX-Lab/CXRGazeLearn}.

\section{Related Works}
Public datasets have catalyzed data-driven advancements in chest X-ray image analysis, utilizing networks like ResNet, DenseNet, and EfficientNet for CXR-based disease classification \cite{ravi2022deep}. Beyond traditional classification approaches, limited recent efforts have explored the integration of radiologists' eye gaze data to incorporate human visual attention into deep learning (DL) frameworks, leveraging this unique modality to guide feature learning for enhanced performance and better understanding of diagnostic criteria. In GazeRadar, Bhattacharya et al. \cite{bhattacharya2022gazeradar} fused radiomics and visual attention features to perform pulmonary disease classification. Later, the same group \cite{bhattacharya2022radiotransformer} proposed RadioTransformer that utilizes visual attention in a cascaded global-focal Transformer framework for CXR classification. Zhu et al. \cite{zhu2022gaze} used the visual saliency maps to guide the class activation mapping to boost diagnostic accuracy. More recently, Wang et al. \cite{wang2023gazegnn} proposed GazeGNN, which uses gaze duration to weigh local image features for CXR diagnosis through a graph neural network. Finally, Zhu et al. \cite{zhu2022multi} proposed a multi-task UNet (MT-UNet) for joint CXR diagnosis and clinical visual saliency map prediction by adopting an elaborate uncertainty-based loss balancing scheme with trainable parameters. To remove the demand for eye gaze data as the input with CXR in clinical deployment, where eye-tracking devices are often not available, in our study, we decided to tackle the challenge with a multi-task, multi-stage cooperative learning strategy to jointly boost the performance of disease diagnosis and visual saliency map prediction further.

\noindent
One critical challenge in multi-task learning is the potential suboptimal performance caused by imbalanced optimization schedules of sub-loss functions, where one task’s loss dominates during training \cite{Vandenhende2020MultiTaskLF}. To tackle this, a few strategies \cite{Kendall2017MultitaskLU,Chen2017GradNormGN,Guo2018,SenerKoltun2018} have been proposed. Chen et al. \cite{Chen2017GradNormGN} proposed GradNorm, a gradient normalization strategy for adaptive loss balance. Guo et al. \cite{Guo2018} leveraged the key performance metric as a measure of task difficulty to dynamically prioritize hard tasks during training. To boost scalability with the dimensionality of the gradients and number of tasks, Sener and Koltun \cite{SenerKoltun2018} employed an upper bound for more efficient multi-objective loss optimization. In the work of Zhu et al. \cite{zhu2022multi}, the framework of Kendall et al. \cite{Kendall2017MultitaskLU}, which uses a learnable coefficient to dynamically balance the different loss functions of different sub-tasks was adopted to ensure effective task performance, and showed excellent results for joint CXR diagnosis and clinical visual saliency map prediction. In contrast, tailored towards the designated tasks, our proposed framework avoids the need for more complex loss balancing methods entirely by adopting a multi-stage cooperative learning strategy, together with a novel dual encoder UNet and multi-scale feature fusion. This approach allows each task to train independently and collaboratively to offer enhanced outcomes in comparison to previous relevant works.

\section{Methods and Materials}
An overview of our proposed technique is illustrated in Fig. 1, which is composed of a dual-encoder residual squeeze-and-excitation UNet (Res\_SE-UNet) to predict visual saliency maps, a DenseNet-201 to encode image features, and a classifier module with multi-scale feature-fusion to provide CXR diagnosis. In our cooperative learning framework, these components were trained in three main stages, in the order of the DenseNet-201 feature encoder, residual squeeze-and-excitation UNet, and multi-scale feature-fusion classifier, to allow a gradual introduction of cooperation of the two learning tasks.

\subsection{Stage 1: DenseNet-201 Feature Encoder}
As DenseNets have shown great performance in image classification tasks, we used the DenseNet-201 as a key feature encoder for the designated tasks. To augment its robustness in feature representation, we first pretrained the network using the CXR data with a contrastive triplet loss \cite{hoffer2015deep}, which minimizes the distances between similar image pairs while maximizing those between dissimilar ones. Then, the pretrained DenseNet201 was finetuned for the task of classifying a CXR scan into normal, pneumonia, or heart failure, with a cross-entropy loss. Here, we used all layers of the DenseNet-201 before the last dense block as our image feature encoder.  

\subsection{Stage 2: Visual Saliency Map Prediction}
The objective of Stage 2 was to generate a visual saliency map for an input CXR scan that mirrors the attention patterns of medical professionals when diagnosing conditions. Following Stage 1, the feature from the trained DenseNet201 encoder (frozen in Stage 2 and 3) was fed into a modified UNet model at the beginning of its decoding path, together with the compressed feature of the same input image from the UNet's encoding branch \cite{ronneberger2015u}. Note that here, we modified the UNet's encoder blocks with Residual and Squeeze-and-Excitation (SE) blocks (see Fig. 1) to enhance its robustness and training stability \cite{liu2019building}. By leveraging features from two distinct image encoders to extract richer and more nuanced information from the CXR images, we intended to enhance the accuracy of visual saliency map prediction at the decoder end. During training, the DenseNet201 feature encoder was kept frozen, while the Res\_SE-UNet was trained using paired CXR images and visual saliency maps, and a Kullback–Leibler (KL) divergence loss.

\subsection{Stage 3: Multi-Scale Feature-Fusion Classifier}
In the final stage, we concatenated the feature from the DenseNet-201 encoder and that from the last upsampling layer of the Res\_SE-UNet, and fed them into a simple neural network classifier (Fig. 1) to conduct CXR diagnosis (normal, pneumonia, or heart failure). This enabled full collaboration of two designated tasks. Here, both the DenseNet-201 encoder and Res\_SE-UNet were frozen, and only the classifier was trained based on a cross-entropy loss. 

\subsection{Dataset and Evaluation Metrics}
In our study, we used the ``chest X-ray dataset with eye-tracking and report dictation" dataset \cite{karargyris2021creation} to develop and validate our proposed algorithm. The dataset contains 1083 chest X-ray scans from the MIMIC-CXR dataset \cite{johnson2019mimic}, with their corresponding diagnostic results (normal, pneumonia, or heart failure), anatomical segmentation, radiologists' audio with dictation, and finally, additional eye-tracking data collected during radiological diagnosis. The eye-tracking data was obtained with a GP3 gaze tracker by Gazepoint (Vancouver, Canada), with an accuracy of around 1° of visual angle at a sampling rate of 60 Hz. For our study, we utilized the visual saliency maps that represent the gaze location and attention over time as a heat map. As the X-ray images have different resolutions and dimensions, we padded and resampled them to the resolution of 640 × 512 pixels, and normalized the pixel intensity to [0,1] per image. 

For the proposed algorithm and comparison methods, we focus on the accuracy evaluation of CXR diagnosis (classification of normal, pneumonia, or heart failure) and visual saliency map prediction. Specifically, for \textbf{diagnostic quality}, we measured the area under the curve (AUC) metric for multi-class classification \cite{fawcett2006introduction} and classification accuracy (ACC) for the overall classification performance, as well as AUCs for three individual classes. In terms of \textbf{visual saliency prediction quality}, we adopted the KL divergence, Pearson’s correlation coefficient (PCC), and histogram similarity (HS), which measure the content and spatial distribution similarity between the predicted and gaze heat maps. While a lower KL divergence is preferred, higher values for the rest of the metrics signify better results. For the metrics that assess the quality of visual saliency map production, two-sided paired sample t-tests were performed to confirm the superiority of our proposed technique against the comparison methods, and a p-value lower than 0.05 was used to declare statistical significance.

\begin{figure}[!htbp]
\centering
\includegraphics[width=1.0\textwidth]{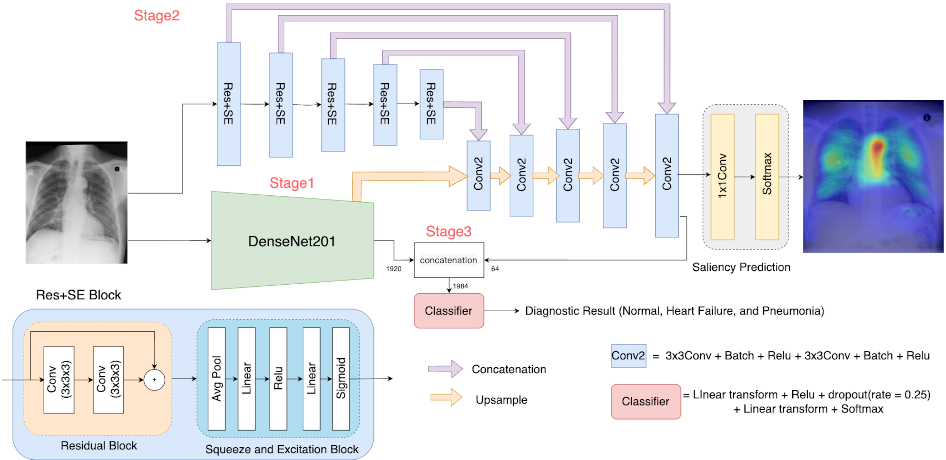}
\caption{\label{fig:final} An overview of the proposed dual-encoder multi-task UNet architecture with multi-stage cooperative learning.}
\end{figure}

\subsection{Experimental Setup and Ablation Studies}
For our experiment, 100 CXR scans, along with the corresponding labels and visual saliency maps, were randomly selected as a test set while the remaining 983 samples were allocated for training purposes \cite{wang2023gazegnn}. Our proposed method involved training at different stages, with the steps detailed in Section 3.1 $\sim$ 3.3. For each stage, the training process is conducted over 50 epochs. We utilized the Adam optimizer with its default parameters \cite{kingma2014adam} (\( \beta_1 = 0.9 \),  \( \beta_2 = 0.999 \), and \( \epsilon = 1 \times 10^{-8} \)) with a learning rate of 0.0001. Early stopping was employed to avoid overfitting, and model training was performed on a desktop computer with Intel Core i9 CPU, Nvidia GeForce RTX 3090 GPU, and 24GB RAM. To validate the performance of our proposed technique, we compared it against the previously established MT-UNet \cite{zhu2022multi}, GazeGNN \cite{wang2023gazegnn}, Inception ResNet v2 \cite{ElAsnaoui2021} (no gaze maps involved), and UNet \cite{ronneberger2015u} for one or both of the target tasks (see Tables 1 \& 2 for details), using the same training setup as our proposed method.

To help gain further insights into the individual elements of our proposed architecture, we conducted a series of four ablation studies. \textbf{First}, to investigate the benefit of adopting Residual and SE blocks in UNet for visual saliency prediction, we compared the heatmap generation quality between only using the Res\_SE-UNet and a standard UNet (UNet\_S), without the boosting from the DenseNet-201. \textbf{Second}, to confirm the contribution of the pretrained DenseNet-201 encoder, we further compared the saliency map prediction accuracy  with and without the encoder module (full network vs. Res\_SE-UNet). \textbf{Third}, the image classification accuracy of the full network vs. DenseNet-201 with contrastive learning was investigated to inspect the benefit of multi-scale feature fusion. \textbf{Finally}, to reveal the advantage of contrastive learning for pretraining the DenseNet-201, we compared the results with and without contrastive learning pretraining for the network in CXR classification.  

\section{Results}
\subsection{Performance of the proposed method}
The quantitative accuracy assessments of CXR classification and saliency map prediction for our proposed dual-encoder feature-fusion UNet and the comparison techniques, including the evaluations for ablation studies are presented in Tables 1 and 2. The arrows within the tables indicate the desired trends of the associated metrics. Overall, our proposed method has achieved an AUC of 0.925 and an accuracy of 80\% for CXR diagnosis, outperforming all the comparison methods, including GazeGNN \cite{wang2023gazegnn}, which ranked as the second best method and uses both the X-ray scan and gaze data as inputs. When looking at the AUC results per class, our method achieved the best score for all class-wise AUCs, except for the category of Normal CXR. In terms of visual saliency prediction, our proposed technique significantly outperformed the MT-UNet \cite{zhu2022multi} and standard UNet (p$<$0.01) in all evaluation metrics while reaching better scores on average over our proposed Res\_SE-UNet (p$>$0.05 for CC, p$<$0.01 for KL and HS). In addition, to qualitatively assess the visual saliency map generation, we present the results from our proposed method, against those produced by the MT-UNet \cite{zhu2022multi}, the gradCAM results \cite{seerala2021grad} from the DenseNet-201 component of our proposed DL architecture, and the ground truths. We can see that our proposed method has a higher resemblance to the ground truths, while the gradCAM outputs have a relatively large discrepancy from the human gaze pattern, which reveals the required scanned areas for full diagnostic criteria.   

\begin{table}[ht]
\centering
\caption{Accuracy assessment of chest X-ray classification for our method, MT-UNet \cite{zhu2022multi}, DenseNet201 with contrastive pretraining (DNet201-CL), DenseNet201 (DNet201), Inception ResNet v2 (IRNetv2)\cite{ElAsnaoui2021}, and GazeGNN \cite{wang2023gazegnn}. The arrows indicate the desired trends of the metric. }
\begin{tabular}{lcccccc}
\hline
Metric & \textbf{Ours} & MT-UNet & DNet201-CL & DNet201 & IRNetv2 & GazeGNN \\
\hline
AUC (Normal) $\uparrow$ & 0.953 & 0.935 & 0.961 & \textbf{0.964} & 0.878 & 0.899 \\
AUC (Heart failure) $\uparrow$ & \textbf{0.927} & 0.881 & 0.865 & 0.897 & 0.873 & 0.881 \\
AUC (Pneumonia) $\uparrow$ & \textbf{0.894} & 0.687 & 0.859 & 0.849 & 0.602 & 0.823 \\
AUC $\uparrow$ & \textbf{0.925} & 0.847 & 0.889 & 0.880 & 0.794 & 0.868 \\
Accuracy $\uparrow$ & \textbf{0.800} & 0.640 & 0.750 & 0.690 & 0.600 & 0.730\\
\hline
\end{tabular}
\label{tab:classification_performance}
\end{table}

\begin{table}[ht]
\centering
\caption{Accuracy assessment of visual saliency map prediction (mean$\pm$std) for our method, MT-UNet \cite{zhu2022multi}, UNet with a modified Residual-Squeeze-and-Excitation encoder (Res\_SE-UNet), and a standard UNet (UNet\_S).}
\setlength{\tabcolsep}{10pt}
\begin{tabular}{lcccc}
\hline
Metric & \textbf{Ours }& MT-UNet & Res\_SE-UNet & UNet\_S \\
\hline
KL $\downarrow$ & \textbf{0.706 $\pm$ 0.183} & 0.747 $\pm$ 0.185 & 0.747 $\pm$ 0.193 & 0.781 $\pm$ 0.202 \\
CC $\uparrow$ & \textbf{0.576 $\pm$ 0.113} & 0.545 $\pm$ 0.109 & 0.560 $\pm$ 0.108 & 0.531 $\pm$ 0.109 \\
HS $\uparrow$ & \textbf{0.552 $\pm$ 0.055} & 0.535 $\pm$ 0.055 & 0.539 $\pm$ 0.058 & 0.527 $\pm$ 0.056 \\
\hline
\end{tabular}
\label{tab:performance_metrics}
\end{table}

\subsection{Ablation studies}
We performed four ablation studies for our proposed framework. \textbf{First}, when comparing the saliency map generation between the Res\_SE-UNet and the standard UNet, we observed a significant improvement in all metrics (p$<$0.05), indicating the positive impact of SE and residual blocks to better capture task-relevant image features. \textbf{Second}, when incorporating the pretrained DenseNet-201 encoder into the Res\_SE-UNet to form the dual-encoder setup (i.e., the full model), the quality of visual saliency prediction was further enhanced, with a large drop of KL divergence from 0.747 to 0.706. \textbf{Third}, in terms of CXR classification, the full model of our proposed technique was compared against the  DenseNet-201 pretrained with contrastive learning (DNet201-CL), and showed great improvement (Accuracy: 80\% vs. 75\%). Here, both the second and third studies showcased boosted performance by leveraging the collaboration of the classification and prediction tasks. \textbf{Lastly}, to confirm the benefit of contrastive pretraining for the DenseNet-201 feature encoder, the CXR classification performance of the network with and without CL pretraining (DNet201-CL vs. DNet201) were compared in Table 1. With the AUC and accuracy increases of 0.009 and 6\%, respectively, the version with CL pretraining is shown to be superior.

\begin{figure}[ht]
\centering
\includegraphics[width=0.9\textwidth]{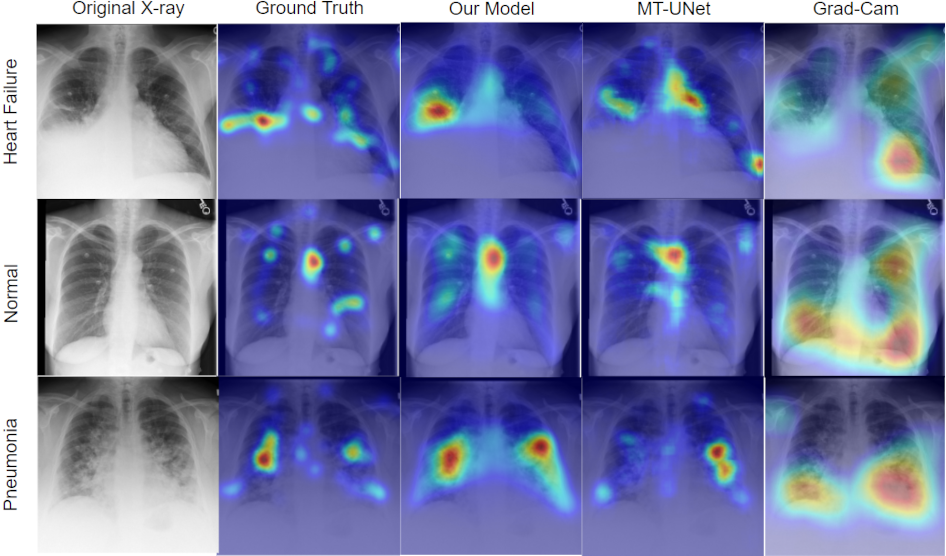}
\caption{\label{fig:final} Qualitative evaluation of visual saliency map prediction across our proposed model and the MT-UNet \cite{zhu2022multi}, against the gradCAM outputs from the DenseNet-201 from our proposed model, the original X-ray scan and the ground truths for exemplary heart failure, normal, and pneumonia cases.}
\vspace{2mm}
\end{figure}

\section{Discussion}

One major feature of our proposed DL framework is to employ a three-stage cooperative learning strategy to address the challenge in multi-task learning, where the loss functions of individual tasks fail to converge at the same rate, resulting in sub-optimal outcomes. Previous approaches tackled this by adopting the uncertainty-based training strategy \cite{zhu2022multi} or independent optimizer that alternates different task-specific losses \cite{pascal2022multi}. These can require more elaborate setups and additional learnable parameters. In contrast, our strategy first used a DenseNet-201 feature encoder pretrained on CXR classification to facilitate visual saliency map prediction, and then further enhance the classification quality by allowing full cooperation of the saliency map and DenseNet-201 features. Our multi-stage training approach also helps reduce computational costs as training each stage is less hardware demanding than training the full model in one go. Based on the ablation studies (see Section 4.2), we showed that such task collaboration with dual-encoder and multi-scale feature-fusion has boosted the performance. In addition, our proposed method outperformed the relatively recent MT-UNet \cite{zhu2022multi} and GazeGNN \cite{wang2023gazegnn}, as well as other baseline techniques in the designated tasks. As in the related reports \cite{wang2023gazegnn, zhu2022multi}, both MT-UNet and GazeGNN have been validated against several state-of-the-art DL models, including ResNets, EfficientNet, Swin Transformers, and VGGSSM \cite{cao2020aggregated}, and showed better results, we decided not to expand the comparison to these models in our study.

In our ablation studies, we confirmed the positive role of contrastive learning with a triplet loss in pretraining the DenseNet-201 feature encoder. Although more recent self-supervised learning techniques \cite{RN46} have provided better results, they could be resource-demanding due to the preference for large batch sizes. We will further explore these methods in the near future. For visual saliency map prediction, we proposed the Res\_SE-UNet, with Residual and SE blocks. While the SE block can increase feature representation by dynamic channel-wise feature recalibration, the residual block facilitates gradient flow during training. This modification was proven to be instrumental in our method. As the gaze attention during diagnosis reflects the positions and importance of local features relevant to a diagnosis, such information may efficiently guide DL-based algorithms for radiological tasks and provide helpful insights for human users (e.g., clinicians in training). Previous investigations \cite{luis2023integrating, wang2022follow, saab2021observational, lanfredi2021comparing} have confirmed the constructive impacts of incorporating gaze saliency maps on improving radiological diagnosis, either as input features with the medical scans or through auxiliary tasks (e.g., regularization of CAM results). This is also well echoed in our study. As shown in Fig. 2, although the CAM has been widely used to provide visual explanation for various DL models, in terms of CXR diagnosis, the CAM pattern does not necessarily coincide with real human attentions and is more ambiguous. Interestingly, in the Heart Failure case of Fig. 2, the human gaze shows attention to the pleural effusion in the right chest, a key diagnostic criteria while the CAM only highlighted the heart. Thus, it would be highly beneficial to employ both visual explanations jointly, and our proposed method conveniently supports this option. The inference time of our model is fast ($\sim$0.3 seconds), which is suitable for clinical use.

In our current study, we utilized the visual saliency map, which is an accumulation of temporally sampled gaze locations to provide a potential explanation for the diagnostic results. However, additional representation of the gaze pattern, such as the scanpaths can also be used to enrich the understanding of the diagnostic procedure and further enhance the accuracy of the diagnostic algorithm to open doors for additional human-machine interaction. Although it is still a challenging task and has not been adopted for clinical scans, ongoing scanpath prediction research in natural images \cite{scandmm2023} is progressing steadily. We will explore venues to incorporate such information in radiological diagnosis in the future.

\section{Conclusion}

We have proposed a novel multi-task DL model using a dual-encoder UNet with multi-scale feature-fusion for CXR diagnosis and visual saliency prediction. The proposed DL model benefits from our multi-stage cooperative learning strategy to best optimize the training of individual tasks, with a gradual introduction of collaboration. The experimental results demonstrate that our method outperforms existing techniques, showcasing its potential to enhance alignment with clinical decision-making processes. Furthermore, the flexibility of our framework suggests its applicability to other domains where human behavior data, such as gaze patterns, can complement deep learning models.

\section*{Acknowledgment}
The study was supported by the Natural Science and Engineering Research Council of Canada. Y. Xiao is supported by supported by the Fond de la Recherche en Santé du Québec (FRQS-chercheur boursier Junior 1) and Parkinson Quebec.

\section*{Conflict Of Interest Statement}
The authors have no relevant conflicts of interest to disclose.

\section*{References}
\addcontentsline{toc}{section}{\numberline{}References}
\vspace{-4mm}

\bibliography{example.bib}      

\bibliographystyle{medphy.bst}

\end{document}